\documentclass[10pt, aps,prc,twocolumn,superscriptaddress,preprintnumbers,
amsmath, 
floatfix,
longbibliography,
nofootinbib
]{revtex4-1}
\usepackage[T1]{fontenc}
\usepackage[utf8x]{inputenc} 
\usepackage{adjustbox}          
\usepackage[caption=false]{subfig}
\usepackage{url}
\usepackage{color}
\usepackage{float}
\usepackage[pdftex,colorlinks=true, linkcolor = blue, citecolor=blue,urlcolor=blue, bookmarksnumbered=true, bookmarksopen=true]{hyperref}
\usepackage{longtable}
\usepackage{amsfonts}
\usepackage{dsfont}
\usepackage{wrapfig,bm} 
\usepackage[normalem]{ulem}
\usepackage{MnSymbol}

\newcommand{\beq}{\begin{equation}}
\newcommand{\eeq}{\end{equation}}
\newcommand{\bea}{\begin{eqnarray}}
\newcommand{\eea}{\end{eqnarray}}

\begin{document}
\title{Large Amplitude Collective Motion and Dissipation in the Ground State and the First Isomeric Wells in the Neutron-Induced Fission of $^{235}$U}

\author{Ibrahim Abdurrahman}
\affiliation{ Facility for Rare Isotope Beams, Michigan State University, East Lansing, Michigan 48824, USA}
 \affiliation{ Theoretical Division, Los Alamos National Laboratory, Los Alamos, New Mexico 87545, USA}

 \author{Matthew Kafker}
 \affiliation{Cyclotron Institute, Texas A\&M University, College Station, Texas 77843, USA}
\affiliation{Department of Physics,%
  University of Washington, Seattle, Washington 98195--1560, USA} 

 \author{Aurel Bulgac }
\affiliation{Department of Physics,%
  University of Washington, Seattle, Washington 98195--1560, USA} 

  \author{Ionel Stetcu}
 \affiliation{ Theoretical Division, Los Alamos National Laboratory, Los Alamos, New Mexico 87545, USA}

\date{\today}

\begin{abstract}
In fission induced by low energy neutrons, the mother nucleus spends a significant fraction of the time in the ground state and isomer wells, eventually passing beyond the outer barrier, where the primary fission fragments properties are defined. Despite this, the dynamics of these two early stages have not been investigated using microscopic models. This study examines the evolution of the mother nucleus in both wells separately, using time-dependent density functional theory, which has been previously used to treat the saddle-to-scission stage of fission for $^{235}$U(n,f) reactions. These two early stages of fission are essential blocks in the final theory of the formation and evolution of a compound nucleus. The present study shows that the dynamics in both wells is strongly dissipative, similar to the dynamics from saddle to scission. It also reveals that while the initial mass asymmetry of the system quickly settles to very small fluctuations in the ground state well, in the isomeric well, the mass asymmetry oscillates with a rather large amplitude, in almost harmonic motion. Furthermore, with low probability, neutrons are emitted in both wells.

\end{abstract} 

\preprint{NT@UW-26-21}
\preprint{LA-UR-26-27006}

\maketitle


Current microscopic treatments of induced fission, primarily time-dependent density functional theory (TDDFT), have focused exclusively on outer saddle-to-scission dynamics~\cite{Bender:2020}. It is in this phase of induced fission where the primary fission fragment (FF) properties are determined. However, during induced fission, the fissioning nuclear system (FNS), which is formed immediately after the neutron is absorbed, $^{235}$U+n $\rightarrow ^{236}$U, spends a very long time until it reaches the outer saddle, much longer than it does evolving from saddle to scission, according to Bohr's insight~\cite{Bohr:1936,Bohr:1936a}.  A microscopic theoretical treatment of the fission dynamics from the moment a neutron is absorbed until it reaches the outer fission barrier is highly desirable. Unfortunately, this task is notoriously difficult, and at the present there exists no microscopic treatment of this entire time-dependent stage of fission, which involves the formation, the evolution, and the eventual decay of the compound nucleus (CN)~\cite{Bohr:1936,Bohr:1936a}. N. Bohr conjectures this time scale is very long, and while, currently, there is no microscopic framework to evaluate it, it's magnitude can be estimated from either nuclear level spacings in neutron capture reactions $^{235}$U(n,$\gamma$) \cite{Bohr:1969,Jandel:2012}, from crystal blocking measurements of the CN lifetime in heavy-ion reactions~\cite{Andersen:2007,Morjean:2008,Back:2020}, or the competition between fission and quasi-fission in heavy-ion reactions~\cite{Toke:1985,Rietz:2013,Hinde:2021}.  All of the aforementioned indicate CN lifetimes of the order of $10^{-18\ldots -16}$ secs. This is significantly longer than the evolution of the system from saddle to scission, which takes on the order of $3- 30\times 10^{-21}$ secs~\cite{Vandenbosch:1973,Gonnenwein:2014,Bulgac:2019c,Bulgac:2020,Bender:2020,Bulgac:2025}. 

 A fully microscopic framework to describe CN lifetimes for fissioning nuclei has not been formulated yet, with only some qualitative semiclassical estimates put forward, at present. There are several significant difficulties in constructing such a theory. First, the tunneling of a nucleus through different barriers, which leads to an increased tunneling time in the case of a dissipative motion~\cite{Caldeira:1983}, has yet to be implemented in TDDFT. One long standing potential implementation, the imaginary time-dependent Hartree-Fock~\cite{Levit:1980,Negele:1989}, or, by extension, an imaginary TDDFT, has never been successfully applied in practice. Second, unlike the FNS formed after neutron absorption, a CN is characterized by a greatly enhanced quantum complexity, involving the extension of the microscopic description with an ensemble of  ${\cal O}(10^4...10^6)$ time-dependent generalized Slater determinants. This is substantially more complex than the single time-dependent generalized Slater determinant involved in a single TDDFT trajectory. Despite the difficulties, recently, it is hopeful such a framework might be realized using an extension of the Generator Coordinate Method (GCM)~\cite{Bulgac:2024d,Kafker:2026}. 

To further illustrate the point: consider a low energy neutron, with kinetic energy, $E_{kin}\leq $ 5 MeV, incident on $^{235}$U.  The resulting system, $^{236}$U, is formed at a low angular momenta state $l=\cal{O}$(1) $\hbar$~\cite{Stetcu:2021}, and has an excitation energy $E_{ex}\approx 6...11$ MeV. The number of quantum states of the incident many-body wave function for n$+^{235}$U reactions is of order $\cal{O}$(1), since $^{235}$U is initially in the ground state.  Within TDDFT this initial state is described by a single (generalized) Slater determinant.\footnote{One can formally argue that after restoring various broken symmetries the ground state of a nucleus is a sum over many Slater determinants, which is a valid statement for a superconductor at $T = 0 $ K. However, a generalized Slater determinant always describes a many body system in a mean field approximation and thus independent quasiparticle motion.} Hence, nuclear fission dynamics when treated within TDDFT, even though strongly dissipative in character, is still mean field dynamics~\cite{Bender:2020,Bulgac:2016,Bulgac:2019c,Bulgac:2020}. It is a quantum mechanical state described by a many-body wave function, which should not be conflated with that corresponding to Bohr's CN, which corresponds to a many-body wave function with significantly more correlated quantum dynamics. 

It is worth noting that although all current implementations of TDDFT applied to nuclear systems are, in some sense, an adiabatic formulation of TDDFT, i.e. local in time, as the detailed discussion in Ref.~\cite{Bender:2020} clarifies, TDDFT still describes some aspects of collective motion dissipation: such as Landau damping and even a quantum implementation the Boltzmann collision integral~\cite{Bulgac:2022} beyond its well known semiclassical extension~\cite{Nordheim:1928,Uehling:1933}. However, the bulk of the mean field fluctuations are not included in either of these extensions, see also Ref.~\cite{Bulgac:2024a}, except in GCM or CI mixing~\cite{Bulgac:2024d,Kafker:2026}. 

Using variants of the Fermi gas model,
\begin{align}
&\rho(U) = \frac{\sqrt{\pi}}{12a^{1/4}U^{5/4}}\exp(2\sqrt{aU}), \label{eq:Fg}\\
&\rho(U) = \frac{\sqrt{\pi}}{12a^{1/4}U^{5/4}}\exp(2\sqrt{a(U-E_{bk})}),\label{eq:bsFg}
\end{align}
where Eq.~(\ref{eq:Fg}, \ref{eq:bsFg}) are the Fermi gas and back-shifted by $E_{bk}$ Fermi gas model, 
or experimental low energy neutron resonances~\cite{Vandenbosch:1973,Bohr:1969,Egidy:2005,Goriely:2008,Jandel:2012,Talou:2021}, 
the level density can be estimated to be $\rho(U)\approx 10^5$/MeV, when $U$ is the excitation energy equal to 
the neutron separation energy of the just formed nuclear system $^{236}$U. The average 
level density in the fission isomer well is lower, as this well is approximately 2-3 MeV higher than the ground state well.
On the way to scission, the nuclear level density of the FNS increases significantly, 
and the energy released, i.e. the total excitation energy (TXE) of 
two fission fragments (FFs) plus the total kinetic energy (TKE) of the FFs, is on average $\approx 200 $ MeV. This is considerably higher excitation energy than the FNS in the ground state well. This illustrates the extremely dissipative nature of fission dynamics, which starts with a nuclear system before neutron absorption $n+^{235}$U, that is characterized by a very simple many-body wave function, while the final state is an extremely complex quantum mechanical state with a very large intrinsic entropy. This is the main idea behind Niels Bohr's conjecture that after $^{235}$U absorbs a low energy neutron the emerging system becomes a compound nucleus~\cite{Bohr:1936}, with a relatively high intrinsic excitation energy, corresponding to a relatively large nuclear level density. Based on this type of logic Bohr postulated that in its final stage the CN retains no memory of it's formation. 

The loss of memory is equivalent to Boltzmann's~\cite{Boltzmann:1872} and Poincar\'e's statements about the evolution of  complex mechanical systems, which are, nevertheless,  governed by time-reversal invariant equations.  Schr\"odinger's equation is also time reversal invariant, and thus the statement that memory is lost should be interpreted with care. Naively, one can assume that within TDDFT framework chaos can arise, since the equations are, by construction, non-linear partial differential equations. However, as shown in Refs.~\cite{Shi:2021,Bulgac:2022c}, the Lyapunov exponents in TDDFT are extremely small and practically negligible. One can conclude that within TDDFT simulations performed with numerical double precision there is no chaotic behavior. 

In the classical limit, the Poincar\'e theorem~\cite{Poincare:1890} relates the loss of memory of the initial conditions to the Poincare recurrence time, the same applies for a CN. To the authors' knowledge so far no microscopic quantum mechanical description for the formation and the evolution of a CN was ever suggested. Published simultaneously with Niels Bohr original paper~\cite{Bohr:1936}, in the editorial~\cite{Bohr:1936a}, a classical model for the CN was suggested, shown in Fig. 1 in Ref.~\cite{Bohr:1936a}, as the collision of a billiard ball with an ensemble of billiard balls at rest in a bowl. The impinging ball quickly looses its initial kinetic energy and the ensemble of billiard balls can lose energy only if that energy is returned eventually to a single billiard ball, which can then ``escape'' from the bowl. This time, according to the Poincar\'e's theorem~\cite{Poincare:1890}, is unimaginably long, in agreement with the main assumption of the classical Boltzmann equation~\cite{Boltzmann:1872}, which implies irreversibility, in apparent disagreement with the time invariance of the classical equations of motion. In the limit of many-body quantum dynamics, this topic is vast and related, in part, to the Eigenstate Thermalization Hypothesis (ETH)~\cite{Neumann:1929,Neumann:2010,Goldstein:2010,Berry:1977,Berry:1991, Deutsch:1991,Srednicki:1994,Rigol:2012,Bulgac:2024,Kafker:2026}, quantum scars~\cite{Heller:1984,Heller:2026}, classical chaos versus quantum chaos within the Gaussian Orthogonal Ensemble~\cite{Bohigas:1984,Bohigas:1993,Mehta:1991,Horoi:1995,Zelevinsky:1996,Haake:2018},  where many similarities and disagreements arise when compared to the classical limit, on which the space here does not allow for further elaboration. The CN starts with a relatively simple many-body wave function and evolves into a many-body wave function characterized by a complexity which is significantly and quite rapidly increasing in time~\cite{Kafker:2026}.         

This work aims to get an insight by examining part of the dynamics by simulating the evolution of the FNS within the ground state and isomer wells separately within TDDFT.  These TDDFT simulations can help to eventually achieve a full microscopic description of the Niels Bohr's CN hypothesis~\cite{Bohr:1936}, which, for fission, has never been demonstrated within a fully microscopic framework. In the future the entire time-dependent formation and evolution of the fissioning nucleus from neutron absorption could, potentially, be attempted using the extended Generator Coordinate Method (eGCM)~\cite{Bulgac:2024d,Kafker:2026}, assuming quantum tunneling can be implemented in a theoretically satisfactory manner. 

\begin{figure*} \includegraphics[width=2.0\columnwidth]{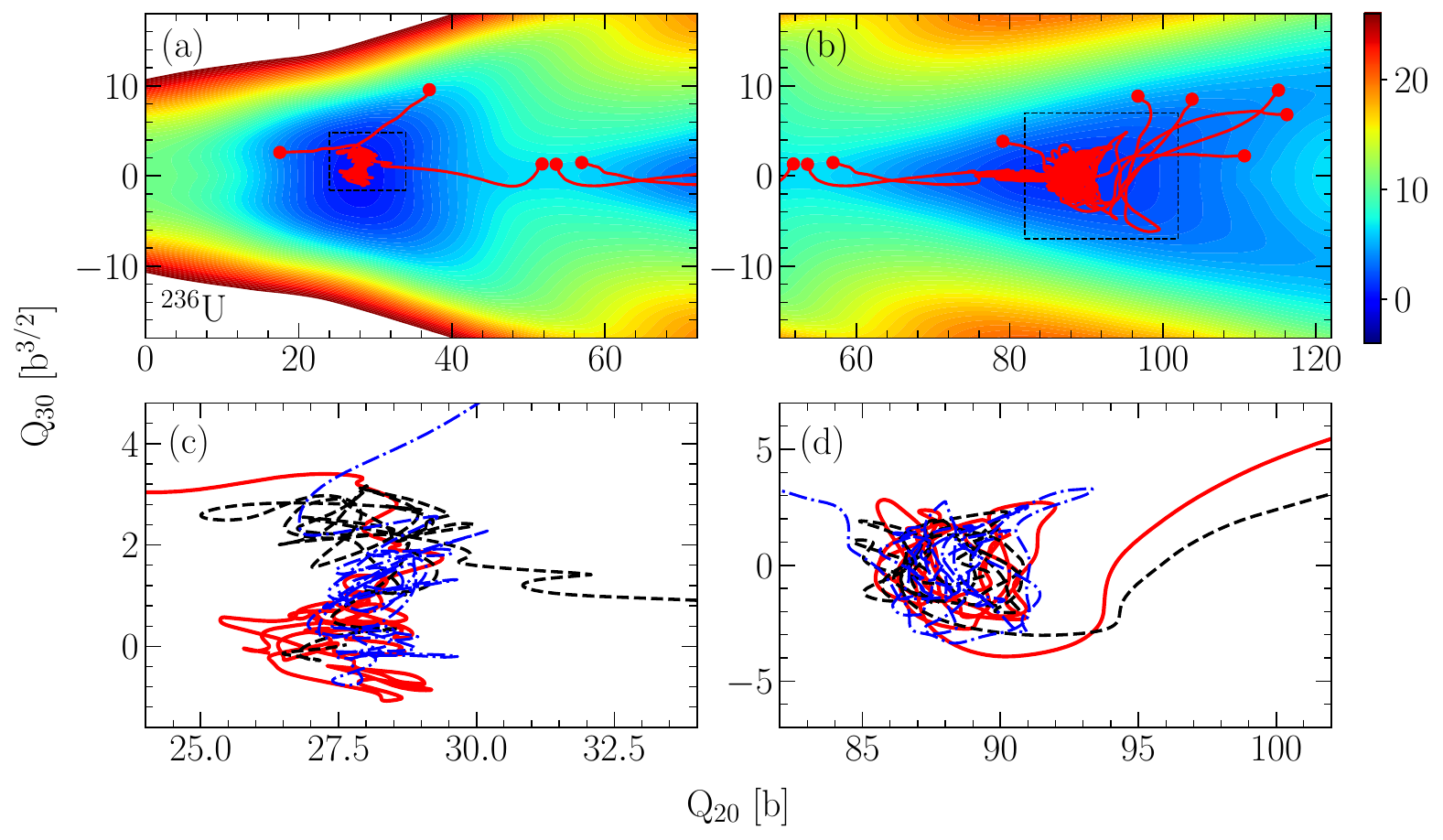}  \caption{ \label{fig:itrajs}  Panels (a) and (b) show the evolution of $^{236}$U on the PES within the ground state and isomer wells respectively. Panel (c) zooms into the ground state well, while panel (d) zooms into the isomer well. The color bar denotes the difference in energy of a point on the PES and the ground state. In panel (d) only three trajectories were plotted.}  \end{figure*}

For this study, several initial conditions of the FNS $^{236}$U, formed after neutron absorption, are prepared at various excitation energies in the ground state and isomer wells of the nuclear potential energy surface (PES), and are shown as red dots in Fig.~\ref{fig:itrajs}. The PES is defined by quadrupole and octupole deformation parameters,
\begin{equation}\label{eqn:deform}
\begin{split}
Q_{20} = \int (2 z'^2 - x'^2 - y'^2) n_t(\bm{r}) d^3r, \\
Q_{30} = \int z'  (2 z'^2 - 3 x'^2 - 3 y'^2) n_t(\bm{r}) d^3r,
\end{split}
\end{equation}
where $n_t(\bm{r})$ is the nucleus' total number density, and $z' = z - z_{\mathrm{cm}}$ (similar for x' and y'). Three trajectories are considered in the ground state well and eight in the isomer well. The FNS is then evolved in time using time-dependent density functional theory extended to superfluid systems~\cite{Shi:2021}. This investigation is restricted to axially symmetric shapes, although it is known triaxiality will play an important role in fission in the region surrounding the first barrier~\cite{Cohen:1974,Ryssens:2015}, and should be included in the future.

Trajectories in the ground state well represent potential configurations of the FNS after the target nucleus has absorbed a neutron. They all start as "cold" nuclei, which are well distinguished by their deformations, that contains the initial imprint of the neutron. As shown in panels (a) and (c) of Fig.~\ref{fig:itrajs}, the trajectories very quickly collapse to the center of the well and their shape parameters undergo a seemingly chaotic evolution in time. This is similar to the trend observed in the saddle to scission evolution of odd systems~\cite{Bulgac:2025}, and ``near-symmetric'' fission~\cite{Abdurrahman:2026}. After a certain time has elapsed, the trajectories become indistinguishable, at least visually, except for their total energies. The isomer well exhibits a similar behavior, as shown in panels (b) and (d) of Fig.~\ref{fig:itrajs}, although there are important differences that will be discussed later.  

\begin{figure} \includegraphics[width=0.99\columnwidth]{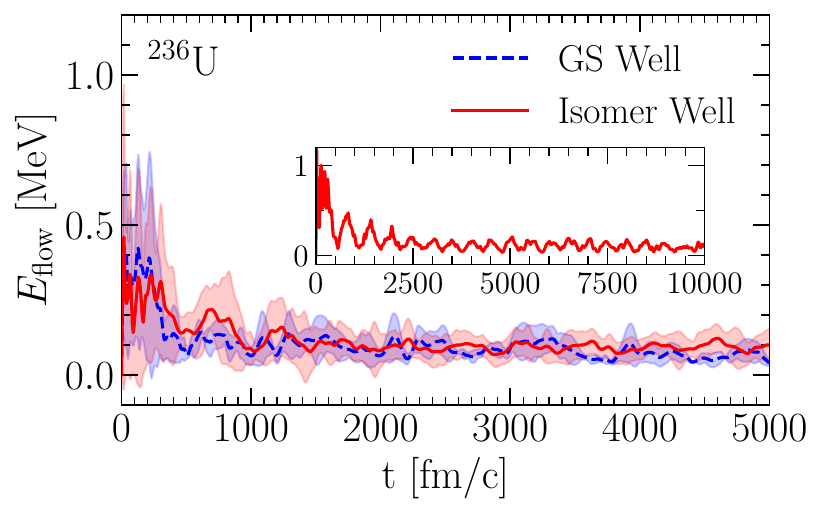}  \caption{ \label{fig:eflow}  The collective flow energy, $E_{\mathrm{flow}} = \sum_{q = n,p} \int d^3 r \frac{\hbar^2}{2m}j^2_q (\bm{r})$, is shown as a function of time for $^{236}$U for trajectories inside the ground state and the isomer wells. The solid lines represent the mean value of the flow energy, and the shaded regions represent one standard deviation away from the mean. The inset shows the flow energy for one trajectory in the isomer well for a longer period of time. }  \end{figure}

The rapid damping of the large amplitude collective motion is caused by strong dissipation at the mean field level. This was previously observed during the saddle to scission dynamics of $^{240}$Pu~\cite{Bulgac:2019c,Bulgac:2020,Bender:2020}, and was also expected to be present in the ground state and fission isomer wells. However, the latter was never directly confirmed, within a fully microscopic framework, until now. As shown in Fig.~\ref{fig:eflow} the collective flow energy starts below 500 keV and quickly decays to a value of $\sim 100$ keV. Equivalently, the FNS starts cold and rapidly thermalizes. The inset of Fig.~\ref{fig:eflow} shows the evolution of the collective flow energy for a longer period of time for one trajectory in the outer well. This suggests that the FNS will remain hot until it approaches the outer saddle~\cite{Bohr:1956,Vandenbosch:1973}, where it will cool again. As a consequence of the strong dissipation, any tunneling process, which is mandatory for spontaneous fission, will be hindered, as demonstrated by~\textcite{Caldeira:1983}. 

\begin{figure} \includegraphics[width=1.0\columnwidth]{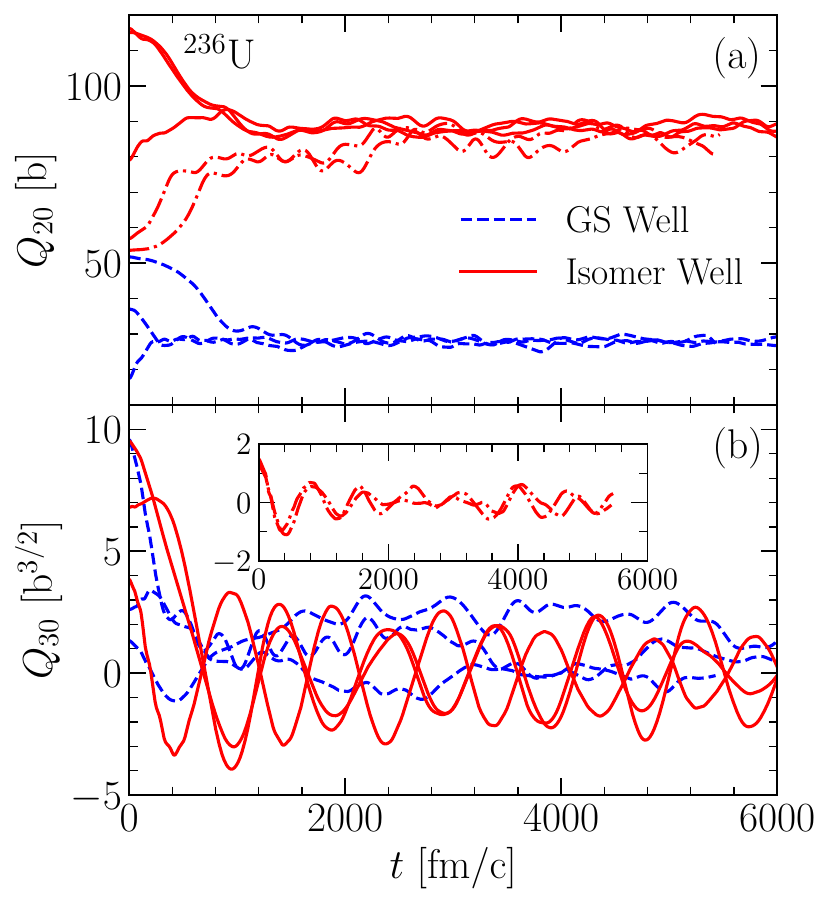}  \caption{ \label{fig:sfluc} Panel (a) shows the FNS's quadrupole moment as a function of time within the ground state and isomer wells. Panel (b) shows the FNS's octupole moment as a function of time within the ground state and isomer wells. Both deformation moments are defined in Eq.(\ref{eqn:deform}). The solid red lines shows three typical trajectories in the isomer well, while the dashed-dotted red lines show two trajectories in the isomer well, that originate near the first barrier.}  \end{figure}

The character of the dynamics of the FNS is further revealed by the evolution of it's shape.  In the top panel of Fig.~\ref{fig:sfluc}, the FNS's quadrupole moment is shown as a function of time. The evolution of this moment is the primary driver for fission. For each well, all trajectories quickly relax to roughly the same central value, and exhibit chaotic fluctuations around it.  The fluctuations are slightly larger in the isomer well than the ground state well. The steady behavior of $Q_{20}$ is consistent with the long time it takes for the FNS to evolve from the ground state well to the outer saddle, a time-scale 6 orders of magnitude longer than its descent from saddle to scission~\cite{Gonnenwein:2014}. 

In panel (b) the evolution of the system's octupole moment is shown. The inset focuses on two trajectories that originate close to the 1st barrier and collapse into the isomer well. These trajectories have a small initial $Q_{30}$. In the ground state well, the evolution of $Q_{30}$ is similar to the evolution of $Q_{20}$, also exhibiting chaotic fluctuations. However, in the isomer well, if the initial value of $Q_{30}$ is large enough, a surprisingly different behavior is observed: $Q_{30}$ oscillates periodically with a fairly large amplitude. The period of the oscillations, $T \approx 960 \pm 70 \mathrm{fm/c}$, is of comparable magnitude to the time it takes the FNS to evolve from saddle to scission~\cite{Bulgac:2019c}, and is roughly equal between trajectories. If the initial value of $Q_{30}$ is below a critical value then it evolves mostly chaotically, even in the isomer well, although some trace of harmonic oscillations is still present.

As the nucleus continues to evolve it will most likely travel through a narrow ``channel'' located at the first barrier~\cite{Bohr:1956,Vandenbosch:1973}. This indicates it will be difficult for the system to acquire a large mass asymmetry as it enters the isomer well. Inside the isomer well the FNS acquires a significant octupole asymmetry~\cite{Ryssens:2015}, even if FNS enters this region with a vanishing octupole moment. As shown here, when $Q_{30}$ exceeds a critical value, it will undergo harmonic oscillations, and the compound's mass asymmetry will then be generated in the isomer well and persist for a long time. If the system never exceeds the critical value of $Q_{30}$, in the isomer well, it will likely fission symmetrically. This implies that the ratio of asymmetric to symmetric fission, which is an experimental observable, can act as a measure of the strength of the fluctuations of $Q_{30}$ in the isomer well, and probe properties of a stage of fission which cannot be measured directly. Tangentially, it should be noted that although the character of the evolution of $Q_{30}$ is distinct in both wells, its maximum value can be quite similar, see Fig.~\ref{fig:sfluc}. 

\begin{figure} \includegraphics[width=1.0\columnwidth]{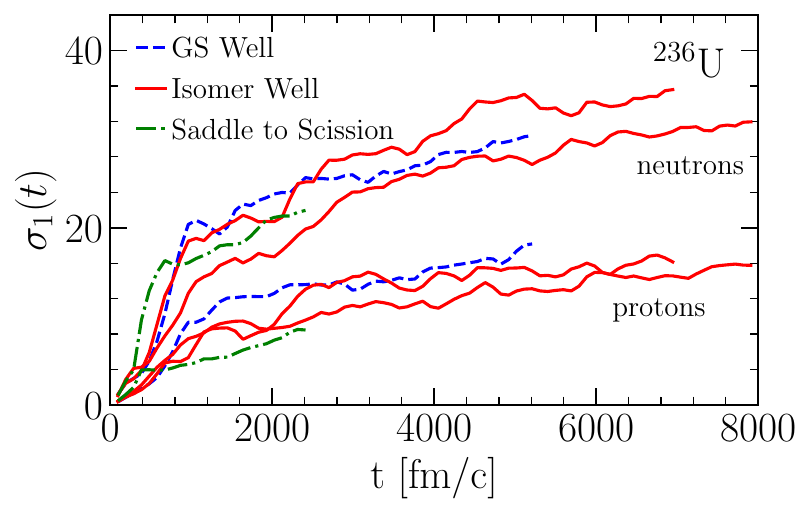}  \caption{ \label{fig:sigma1} The quantity $\sigma_1(t)$, given by Eq.(\ref{eqn:sigma}), is shown for one trajectory in the ground state well, two in the isomer well, and one saddle to scission trajectory, as a function of time. The saddle to scission trajectory starts at the top of the saddle point, and ends when the FFs are separated by 30 fm. The top four lines denote neutrons, while the bottom four lines denote protons.}  \end{figure}

In Fig.~\ref{fig:sigma1}, the compound dynamics in both wells are examined through a different lens: the quantity,
44 \begin{equation} \label{eqn:sigma}
    \sigma_1 (t) = \sum_k|n_k(t)-n_k(0)|
\end{equation}
which characterizes the redistribution of the single particle occupation numbers. A single saddle to scission trajectory was also added for comparison. As was previously seen, the evolution from saddle to scission is non-Markovian, as $\sigma_1(t)$ depends roughly linearly on time~\cite{Bulgac:2024a}. The total single-particle occupation probability $\sum_kn_k(0)$ changes by about $\sigma_1(t)/\sum_k n_k(0)\approx 20 - 25\%$ during the time evolution. 

\begin{figure} \includegraphics[width=1.0\columnwidth]{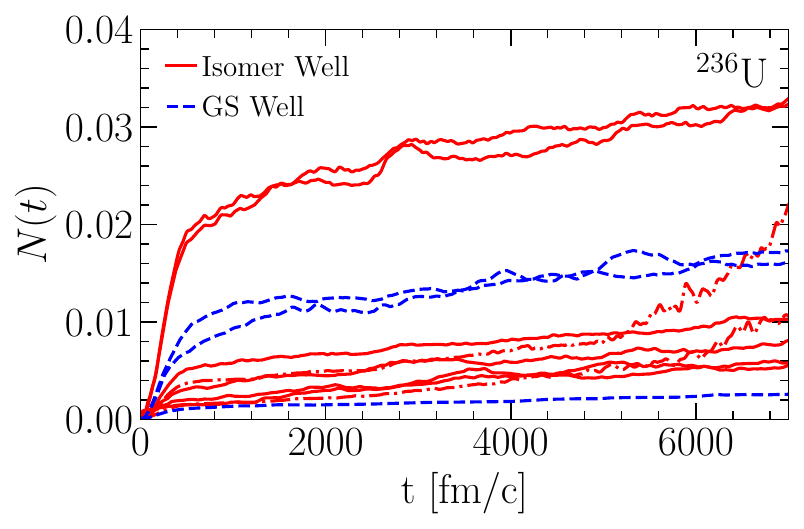}
\caption{ \label{fig:neutrons} The number of neutrons emitted as a function of time are shown for all trajectories in the ground state and isomer wells. The solid red lines shows typical trajectories in the isomer well, while the dashed-dotted red lines show two trajectories in the isomer well, that originate near the first barrier. The number of neutrons are obtained by integrating the neutron density located 17 fm, or more, beyond the center of mass of the FNS along the fission axis. The numbers estimated in this manner are likely to be slightly conservative. }
\end{figure}

Fig.~\ref{fig:neutrons} shows the time-dependence of the number of emitted neutrons while the system is either in the ground or the first isomer wells. In TDDFT this manifests as clouds of neutron density escaping within a fixed volume contained within the total simulation box: $30^2\times60$ fm$^3$. There are two regimes; the first is a rapid jump at early times, likely corresponding to the removal of the deformation constraints at $t=0$ fm/c; the second is a roughly linear emission of neutrons. One possible interpretation is that these neutrons arise from the scattering of the neutrons on the moving walls of FNS $^{236}$U. Similar neutron emission has been previously observed during the saddle to scission dynamics~\cite{Bulgac:2019c} and the neck rupture~\cite{Abdurrahman:2024}. Two trajectories, those which start close to the ground state well but decay into the isomer well, exhibit a second increase of neutrons whose origin is unclear. 

In summary, this study focused on the FNS dynamics in the ground state and isomer wells, which have not been treated microscopically as  far as the authors are aware, despite many open questions. For example, how long does the FNS spend in the ground state well versus the isomer well? It was demonstrated that the dynamics of the FNS in both wells are dissipative, similar to the fission dynamics beyond the outer fission barrier. This result was expected, but never explicitly demonstrated within a fully microscopic framework. Second, the study shows that the initial deformation of $^{236}$U is quickly forgotten, as all trajectories quickly collapse into the ground state well, and become almost indistinguishable from each other, outside of their total energy. In the isomer well, some presence of the initial deformation remains, as mass asymmetry of the FNS generated here persists for a long time, if it exceeds an initial critical value, and is harmonic in character. This is a surprising difference between dynamics in both wells. Furthermore, the strength of the fluctuation of $Q_{30}$ maybe characterized by the ratio of asymmetric to symmetric fission events allowing for a unique experimental probe into an otherwise inaccessible stage of fission. These results further confirm the non-Markovian character of fission, as well as show a potential signal of elastic and/or inelastic neutron scattering. It is important to note, current conclusions are based on a treatment of the fission dynamics at the mean field level and ought to be revisited after beyond mean field fluctuations are added using eGCM~\cite{Bulgac:2024d,Kafker:2026} or potentially other microscopic frameworks.

\section*{Acknowledgments}\label{sec:ack}

I.A. and I.S. were supported by the U.S.
Department of Energy through the Los Alamos National
Laboratory. The Los Alamos National Laboratory is operated
by Triad National Security, LLC, for the National Nuclear
Security Administration of the U.S. Department of
Energy Contract No. 89233218CNA000001. I.A. and I.S. gratefully acknowledge partial support and computational
resources provided by the Advanced Simulation and
Computing (ASC) Program. M.K. was supported by NNSA cooperative Agreement DE-NA0003841. A.B. was supported by the Office of Science, Grant No. DE-FG02-97ER41014  
and partially by NNSA cooperative Agreement DE-NA0003841. This research used resources of the Oak Ridge Leadership Computing 
Facility, which is a U.S. DOE Office of
Science User Facility supported under Contract No. DE-AC05-00OR22725. 

\newpage


\providecommand{\selectlanguage}[1]{}
\renewcommand{\selectlanguage}[1]{}

\bibliography{local_fission}

\begin{thebibliography}{56}%
\makeatletter
\providecommand \@ifxundefined [1]{%
 \@ifx{#1\undefined}
}%
\providecommand \@ifnum [1]{%
 \ifnum #1\expandafter \@firstoftwo
 \else \expandafter \@secondoftwo
 \fi
}%
\providecommand \@ifx [1]{%
 \ifx #1\expandafter \@firstoftwo
 \else \expandafter \@secondoftwo
 \fi
}%
\providecommand \natexlab [1]{#1}%
\providecommand \enquote  [1]{``#1''}%
\providecommand \bibnamefont  [1]{#1}%
\providecommand \bibfnamefont [1]{#1}%
\providecommand \citenamefont [1]{#1}%
\providecommand \href@noop [0]{\@secondoftwo}%
\providecommand \href [0]{\begingroup \@sanitize@url \@href}%
\providecommand \@href[1]{\@@startlink{#1}\@@href}%
\providecommand \@@href[1]{\endgroup#1\@@endlink}%
\providecommand \@sanitize@url [0]{\catcode `\\12\catcode `\$12\catcode
  `\&12\catcode `\#12\catcode `\^12\catcode `\_12\catcode `\%12\relax}%
\providecommand \@@startlink[1]{}%
\providecommand \@@endlink[0]{}%
\providecommand \url  [0]{\begingroup\@sanitize@url \@url }%
\providecommand \@url [1]{\endgroup\@href {#1}{\urlprefix }}%
\providecommand \urlprefix  [0]{URL }%
\providecommand \Eprint [0]{\href }%
\providecommand \doibase [0]{http://dx.doi.org/}%
\providecommand \selectlanguage [0]{\@gobble}%
\providecommand \bibinfo  [0]{\@secondoftwo}%
\providecommand \bibfield  [0]{\@secondoftwo}%
\providecommand \translation [1]{[#1]}%
\providecommand \BibitemOpen [0]{}%
\providecommand \bibitemStop [0]{}%
\providecommand \bibitemNoStop [0]{.\EOS\space}%
\providecommand \EOS [0]{\spacefactor3000\relax}%
\providecommand \BibitemShut  [1]{\csname bibitem#1\endcsname}%
\let\auto@bib@innerbib\@empty
\bibitem [{\citenamefont {Bender}\ and\ \citenamefont {{\it et
  al.}}(2020)}]{Bender:2020}%
  \BibitemOpen
  \bibfield  {author} {\bibinfo {author} {\bibfnamefont {M.}~\bibnamefont
  {Bender}}\ and\ \bibinfo {author} {\bibnamefont {{\it et al.}}},\ }\bibfield
  {title} {\enquote {\bibinfo {title} {Future of nuclear fission theory},}\
  }\href {\doibase 10.1088/1361-6471/abab4f} {\bibfield  {journal} {\bibinfo
  {journal} {J. Phys. G: Nucl. Part. Phys.}\ }\textbf {\bibinfo {volume}
  {47}},\ \bibinfo {pages} {113002} (\bibinfo {year} {2020})}\BibitemShut
  {NoStop}%
\bibitem [{\citenamefont {Bohr}(1936)}]{Bohr:1936}%
  \BibitemOpen
  \bibfield  {author} {\bibinfo {author} {\bibfnamefont {N.}~\bibnamefont
  {Bohr}},\ }\bibfield  {title} {\enquote {\bibinfo {title} {{Neutron Capture
  and Nuclear Constitution}},}\ }\href {\doibase 10.1038/137344a0} {\bibfield
  {journal} {\bibinfo  {journal} {Nature}\ }\textbf {\bibinfo {volume} {137}},\
  \bibinfo {pages} {344 and 351} (\bibinfo {year} {1936})}\BibitemShut
  {NoStop}%
\bibitem [{\citenamefont {{Editorial}}(1936)}]{Bohr:1936a}%
  \BibitemOpen
  \bibfield  {author} {\bibinfo {author} {\bibnamefont {{Editorial}}},\
  }\bibfield  {title} {\enquote {\bibinfo {title} {{Neutron Capture and Nuclear
  Constitution}},}\ }\href {\doibase 10.1038/137351a0} {\bibfield  {journal}
  {\bibinfo  {journal} {Nature}\ }\textbf {\bibinfo {volume} {137}},\ \bibinfo
  {pages} {351} (\bibinfo {year} {1936})}\BibitemShut {NoStop}%
\bibitem [{\citenamefont {Bohr}\ and\ \citenamefont
  {Mottelson}(1969)}]{Bohr:1969}%
  \BibitemOpen
  \bibfield  {author} {\bibinfo {author} {\bibfnamefont {A.}~\bibnamefont
  {Bohr}}\ and\ \bibinfo {author} {\bibfnamefont {B.~R.}\ \bibnamefont
  {Mottelson}},\ }\href@noop {} {\emph {\bibinfo {title} {Nuclear
  Structure}}},\ Vol.~\bibinfo {volume} {I}\ (\bibinfo  {publisher} {Benjamin
  Inc.},\ \bibinfo {address} {New York},\ \bibinfo {year} {1969})\BibitemShut
  {NoStop}%
\bibitem [{\citenamefont {Jandel}\ and\ \citenamefont {{\it
  others}}(2012)}]{Jandel:2012}%
  \BibitemOpen
  \bibfield  {author} {\bibinfo {author} {\bibfnamefont {M.}~\bibnamefont
  {Jandel}}\ and\ \bibinfo {author} {\bibnamefont {{\it others}}},\ }\bibfield
  {title} {\enquote {\bibinfo {title} {New precision measurements of the
  $^{235}\mathbf{U}(n,\ensuremath{\gamma})$ cross section},}\ }\href {\doibase
  10.1103/PhysRevLett.109.202506} {\bibfield  {journal} {\bibinfo  {journal}
  {Phys. Rev. Lett.}\ }\textbf {\bibinfo {volume} {109}},\ \bibinfo {pages}
  {202506} (\bibinfo {year} {2012})}\BibitemShut {NoStop}%
\bibitem [{\citenamefont {Andersen}\ and\ \citenamefont {{\it
  others}}(2007)}]{Andersen:2007}%
  \BibitemOpen
  \bibfield  {author} {\bibinfo {author} {\bibfnamefont {J.~U.}\ \bibnamefont
  {Andersen}}\ and\ \bibinfo {author} {\bibnamefont {{\it others}}},\
  }\bibfield  {title} {\enquote {\bibinfo {title} {{Crystal Blocking
  Measurements of the Time Delay of Fission Induced by $^{32}\mathrm{S}$,
  $^{48}\mathrm{Ti}$, and $^{58}\mathrm{Ni}$ Bombardment of W}},}\ }\href
  {\doibase 10.1103/PhysRevLett.99.162502} {\bibfield  {journal} {\bibinfo
  {journal} {Phys. Rev. Lett.}\ }\textbf {\bibinfo {volume} {99}},\ \bibinfo
  {pages} {162502} (\bibinfo {year} {2007})}\BibitemShut {NoStop}%
\bibitem [{\citenamefont {Morjean}\ and\ \citenamefont {{\it
  others}}(2008)}]{Morjean:2008}%
  \BibitemOpen
  \bibfield  {author} {\bibinfo {author} {\bibfnamefont {M.}~\bibnamefont
  {Morjean}}\ and\ \bibinfo {author} {\bibnamefont {{\it others}}},\ }\bibfield
   {title} {\enquote {\bibinfo {title} {Fission time measurements: A new probe
  into superheavy element stability},}\ }\href {\doibase
  10.1103/PhysRevLett.101.072701} {\bibfield  {journal} {\bibinfo  {journal}
  {Phys. Rev. Lett.}\ }\textbf {\bibinfo {volume} {101}},\ \bibinfo {pages}
  {072701} (\bibinfo {year} {2008})}\BibitemShut {NoStop}%
\bibitem [{\citenamefont {Back}(2020)}]{Back:2020}%
  \BibitemOpen
  \bibfield  {author} {\bibinfo {author} {\bibfnamefont {B.~B.}\ \bibnamefont
  {Back}},\ }\bibfield  {title} {\enquote {\bibinfo {title}
  {xparimentalmeasures of fission time scales},}\ }\href {\doibase
  10.7566/JPSCP.32.010002} {\bibfield  {journal} {\bibinfo  {journal} {JPS
  Conf. Proc.}\ }\textbf {\bibinfo {volume} {32}},\ \bibinfo {pages} {010002}
  (\bibinfo {year} {2020})}\BibitemShut {NoStop}%
\bibitem [{\citenamefont {T{\=o}ke}\ and\ \citenamefont {{\it
  others}}(1985)}]{Toke:1985}%
  \BibitemOpen
  \bibfield  {author} {\bibinfo {author} {\bibfnamefont {J.}~\bibnamefont
  {T{\=o}ke}}\ and\ \bibinfo {author} {\bibnamefont {{\it others}}},\
  }\bibfield  {title} {\enquote {\bibinfo {title} {Quasi-fission --- the
  mass-drift mode in heavy-ion reactions},}\ }\href {\doibase
  https://doi.org/10.1016/0375-9474(85)90344-6} {\bibfield  {journal} {\bibinfo
   {journal} {Nucl. Phys. A}\ }\textbf {\bibinfo {volume} {440}},\ \bibinfo
  {pages} {327} (\bibinfo {year} {1985})}\BibitemShut {NoStop}%
\bibitem [{\citenamefont {du~Rietz}\ and\ \citenamefont {{\it
  others}}(2013)}]{Rietz:2013}%
  \BibitemOpen
  \bibfield  {author} {\bibinfo {author} {\bibfnamefont {R.}~\bibnamefont
  {du~Rietz}}\ and\ \bibinfo {author} {\bibnamefont {{\it others}}},\
  }\bibfield  {title} {\enquote {\bibinfo {title} {Mapping quasifission
  characteristics and timescales in heavy element formation reactions},}\
  }\href {\doibase 10.1103/PhysRevC.88.054618} {\bibfield  {journal} {\bibinfo
  {journal} {Phys. Rev. C}\ }\textbf {\bibinfo {volume} {88}},\ \bibinfo
  {pages} {054618} (\bibinfo {year} {2013})}\BibitemShut {NoStop}%
\bibitem [{\citenamefont {Hinde}\ \emph {et~al.}(2021)\citenamefont {Hinde},
  \citenamefont {Dasgupta},\ and\ \citenamefont {Simpson}}]{Hinde:2021}%
  \BibitemOpen
  \bibfield  {author} {\bibinfo {author} {\bibfnamefont {D.J.}\ \bibnamefont
  {Hinde}}, \bibinfo {author} {\bibfnamefont {M.}~\bibnamefont {Dasgupta}}, \
  and\ \bibinfo {author} {\bibfnamefont {E.C.}\ \bibnamefont {Simpson}},\
  }\bibfield  {title} {\enquote {\bibinfo {title} {Experimental studies of the
  competition between fusion and quasifission in the formation of heavy and
  superheavy nuclei},}\ }\href {\doibase
  https://doi.org/10.1016/j.ppnp.2021.103856} {\bibfield  {journal} {\bibinfo
  {journal} {Prog. Part. Nucl. Phys.}\ }\textbf {\bibinfo {volume} {118}},\
  \bibinfo {pages} {103856} (\bibinfo {year} {2021})}\BibitemShut {NoStop}%
\bibitem [{\citenamefont {Vandenbosch}\ and\ \citenamefont
  {Huizenga}(1973)}]{Vandenbosch:1973}%
  \BibitemOpen
  \bibfield  {author} {\bibinfo {author} {\bibfnamefont {R.}~\bibnamefont
  {Vandenbosch}}\ and\ \bibinfo {author} {\bibfnamefont {J.~R.}\ \bibnamefont
  {Huizenga}},\ }\bibfield  {title} {\enquote {\bibinfo {title} {{Nuclear
  Fission}},}\ }\href@noop {} {\bibfield  {journal} {\bibinfo  {journal}
  {Academic Press, New York}\ } (\bibinfo {year} {1973})}\BibitemShut {NoStop}%
\bibitem [{\citenamefont {G{\"o}nnenwein}(2014)}]{Gonnenwein:2014}%
  \BibitemOpen
  \bibfield  {author} {\bibinfo {author} {\bibfnamefont {F.}~\bibnamefont
  {G{\"o}nnenwein}},\ }\href
  {http://depni.sinp.msu.ru/~kuznets/fission/Goennenwein.pdf} {\enquote
  {\bibinfo {title} {Neutron and gamma emission in fission},}\ }\bibinfo
  {howpublished} {LANL Fiesta 2014 Lectures} (\bibinfo {year}
  {2014})\BibitemShut {NoStop}%
\bibitem [{\citenamefont {Bulgac}\ \emph {et~al.}(2019)\citenamefont {Bulgac},
  \citenamefont {Jin}, \citenamefont {Roche}, \citenamefont {Schunck},\ and\
  \citenamefont {Stetcu}}]{Bulgac:2019c}%
  \BibitemOpen
  \bibfield  {author} {\bibinfo {author} {\bibfnamefont {A.}~\bibnamefont
  {Bulgac}}, \bibinfo {author} {\bibfnamefont {S.}~\bibnamefont {Jin}},
  \bibinfo {author} {\bibfnamefont {K.~J.}\ \bibnamefont {Roche}}, \bibinfo
  {author} {\bibfnamefont {N.}~\bibnamefont {Schunck}}, \ and\ \bibinfo
  {author} {\bibfnamefont {I.}~\bibnamefont {Stetcu}},\ }\bibfield  {title}
  {\enquote {\bibinfo {title} {Fission dynamics of $^{240}\mathrm{Pu}$ from
  saddle to scission and beyond},}\ }\href {\doibase
  10.1103/PhysRevC.100.034615} {\bibfield  {journal} {\bibinfo  {journal}
  {Phys. Rev. C}\ }\textbf {\bibinfo {volume} {100}},\ \bibinfo {pages}
  {034615} (\bibinfo {year} {2019})}\BibitemShut {NoStop}%
\bibitem [{\citenamefont {Bulgac}\ \emph {et~al.}(2020)\citenamefont {Bulgac},
  \citenamefont {Jin},\ and\ \citenamefont {Stetcu}}]{Bulgac:2020}%
  \BibitemOpen
  \bibfield  {author} {\bibinfo {author} {\bibfnamefont {A.}~\bibnamefont
  {Bulgac}}, \bibinfo {author} {\bibfnamefont {S.}~\bibnamefont {Jin}}, \ and\
  \bibinfo {author} {\bibfnamefont {I.}~\bibnamefont {Stetcu}},\ }\bibfield
  {title} {\enquote {\bibinfo {title} {{Nuclear Fission Dynamics: Past,
  Present, Needs, and Future}},}\ }\href {\doibase 10.3389/fphy.2020.00063}
  {\bibfield  {journal} {\bibinfo  {journal} {{Frontiers in Physics}}\ }\textbf
  {\bibinfo {volume} {8}},\ \bibinfo {pages} {63} (\bibinfo {year}
  {2020})}\BibitemShut {NoStop}%
\bibitem [{\citenamefont {Bulgac}\ \emph {et~al.}(2025)\citenamefont {Bulgac},
  \citenamefont {Abdurrahman}, \citenamefont {Kafker},\ and\ \citenamefont
  {Stetcu}}]{Bulgac:2025}%
  \BibitemOpen
  \bibfield  {author} {\bibinfo {author} {\bibfnamefont {A.}~\bibnamefont
  {Bulgac}}, \bibinfo {author} {\bibfnamefont {I.}~\bibnamefont {Abdurrahman}},
  \bibinfo {author} {\bibfnamefont {M.}~\bibnamefont {Kafker}}, \ and\ \bibinfo
  {author} {\bibfnamefont {I.}~\bibnamefont {Stetcu}},\ }\bibfield  {title}
  {\enquote {\bibinfo {title} {{Time-Dependent Density Functional Theory
  Description of $^{238}$U(n,f), $^{240, 242}$Pu(n,f), and $^{237}$Np(n,f)
  Reactions}},}\ }\href {\doibase https://doi.org/10.1103/2k8k-vpng} {\bibfield
   {journal} {\bibinfo  {journal} {Phys. Rev. Lett.}\ }\textbf {\bibinfo
  {volume} {135}},\ \bibinfo {pages} {062501} (\bibinfo {year}
  {2025})}\BibitemShut {NoStop}%
\bibitem [{\citenamefont {Caldeira}\ and\ \citenamefont
  {Leggett}(1983)}]{Caldeira:1983}%
  \BibitemOpen
  \bibfield  {author} {\bibinfo {author} {\bibfnamefont {A.O}\ \bibnamefont
  {Caldeira}}\ and\ \bibinfo {author} {\bibfnamefont {A.J}\ \bibnamefont
  {Leggett}},\ }\bibfield  {title} {\enquote {\bibinfo {title} {Quantum
  tunnelling in a dissipative system},}\ }\href {\doibase
  https://doi.org/10.1016/0003-4916(83)90202-6} {\bibfield  {journal} {\bibinfo
   {journal} {Ann. Phys.}\ }\textbf {\bibinfo {volume} {149}},\ \bibinfo
  {pages} {374} (\bibinfo {year} {1983})}\BibitemShut {NoStop}%
\bibitem [{\citenamefont {Levit}\ \emph {et~al.}(1980)\citenamefont {Levit},
  \citenamefont {Negele},\ and\ \citenamefont {Paltiel}}]{Levit:1980}%
  \BibitemOpen
  \bibfield  {author} {\bibinfo {author} {\bibfnamefont {S.}~\bibnamefont
  {Levit}}, \bibinfo {author} {\bibfnamefont {J.~W.}\ \bibnamefont {Negele}}, \
  and\ \bibinfo {author} {\bibfnamefont {Z.}~\bibnamefont {Paltiel}},\
  }\bibfield  {title} {\enquote {\bibinfo {title} {Barrier penetration and
  spontaneous fission in the time-dependent mean-field approximation},}\ }\href
  {\doibase 10.1103/PhysRevC.22.1979} {\bibfield  {journal} {\bibinfo
  {journal} {Phys. Rev. C}\ }\textbf {\bibinfo {volume} {22}},\ \bibinfo
  {pages} {1979} (\bibinfo {year} {1980})}\BibitemShut {NoStop}%
\bibitem [{\citenamefont {Negele}()}]{Negele:1989}%
  \BibitemOpen
  \bibfield  {author} {\bibinfo {author} {\bibfnamefont {J.~W.}\ \bibnamefont
  {Negele}},\ }\bibfield  {title} {\enquote {\bibinfo {title} {{Micrscopic
  Theory of Fission Dynamics}},}\ }\href@noop {} {\bibfield  {journal}
  {\bibinfo  {journal} {Nucl. Phys. A}\ }\textbf {\bibinfo {volume} {502}},\
  \bibinfo {pages} {371c}}\BibitemShut {NoStop}%
\bibitem [{\citenamefont {Bulgac}()}]{Bulgac:2024d}%
  \BibitemOpen
  \bibfield  {author} {\bibinfo {author} {\bibfnamefont {A.}~\bibnamefont
  {Bulgac}},\ }\href@noop {} {\enquote {\bibinfo {title} {{A critical
  assessment of the current implementations of the Generator Coordinate Method,
  Phys. Rev. C (2026) in press}},}\ }\Eprint {http://arxiv.org/abs/2408.02173}
  {arXiv:2408.02173 [nucl-th]} \BibitemShut {NoStop}%
\bibitem [{\citenamefont {Kafker}\ and\ \citenamefont
  {Bulgac}(2026)}]{Kafker:2026}%
  \BibitemOpen
  \bibfield  {author} {\bibinfo {author} {\bibfnamefont {M.}~\bibnamefont
  {Kafker}}\ and\ \bibinfo {author} {\bibfnamefont {A.}~\bibnamefont
  {Bulgac}},\ }\href@noop {} {\enquote {\bibinfo {title} {{Multi-Nucleon
  Transfer Reactions and the Creation and the Evolution of the Compound
  Nucleus}},}\ } (\bibinfo {year} {2026}),\ \Eprint
  {http://arxiv.org/abs/2604.21845} {arXiv:2604.21845 [nucl-th]} \BibitemShut
  {NoStop}%
\bibitem [{\citenamefont {Stetcu}\ \emph {et~al.}(2021)\citenamefont {Stetcu},
  \citenamefont {Lovell}, \citenamefont {Talou}, \citenamefont {Kawano},
  \citenamefont {Marin}, \citenamefont {Pozzi},\ and\ \citenamefont
  {Bulgac}}]{Stetcu:2021}%
  \BibitemOpen
  \bibfield  {author} {\bibinfo {author} {\bibfnamefont {I.}~\bibnamefont
  {Stetcu}}, \bibinfo {author} {\bibfnamefont {A.~E.}\ \bibnamefont {Lovell}},
  \bibinfo {author} {\bibfnamefont {P.}~\bibnamefont {Talou}}, \bibinfo
  {author} {\bibfnamefont {T.}~\bibnamefont {Kawano}}, \bibinfo {author}
  {\bibfnamefont {S.}~\bibnamefont {Marin}}, \bibinfo {author} {\bibfnamefont
  {S.~A.}\ \bibnamefont {Pozzi}}, \ and\ \bibinfo {author} {\bibfnamefont
  {A.}~\bibnamefont {Bulgac}},\ }\bibfield  {title} {\enquote {\bibinfo {title}
  {{Angular Momentum Removal by Neutron and $\ensuremath{\gamma}$-Ray Emissions
  during Fission Fragment Decays}},}\ }\href {\doibase
  10.1103/PhysRevLett.127.222502} {\bibfield  {journal} {\bibinfo  {journal}
  {Phys. Rev. Lett.}\ }\textbf {\bibinfo {volume} {127}},\ \bibinfo {pages}
  {222502} (\bibinfo {year} {2021})}\BibitemShut {NoStop}%
\bibitem [{\citenamefont {Bulgac}\ \emph {et~al.}(2016)\citenamefont {Bulgac},
  \citenamefont {Magierski}, \citenamefont {Roche},\ and\ \citenamefont
  {Stetcu}}]{Bulgac:2016}%
  \BibitemOpen
  \bibfield  {author} {\bibinfo {author} {\bibfnamefont {A.}~\bibnamefont
  {Bulgac}}, \bibinfo {author} {\bibfnamefont {P.}~\bibnamefont {Magierski}},
  \bibinfo {author} {\bibfnamefont {K.~J.}\ \bibnamefont {Roche}}, \ and\
  \bibinfo {author} {\bibfnamefont {I.}~\bibnamefont {Stetcu}},\ }\bibfield
  {title} {\enquote {\bibinfo {title} {{Induced Fission of $^{240}\mathrm{Pu}$
  within a Real-Time Microscopic Framework}},}\ }\href {\doibase
  10.1103/PhysRevLett.116.122504} {\bibfield  {journal} {\bibinfo  {journal}
  {Phys. Rev. Lett.}\ }\textbf {\bibinfo {volume} {116}},\ \bibinfo {pages}
  {122504} (\bibinfo {year} {2016})}\BibitemShut {NoStop}%
\bibitem [{\citenamefont {Bulgac}(2022)}]{Bulgac:2022}%
  \BibitemOpen
  \bibfield  {author} {\bibinfo {author} {\bibfnamefont {A.}~\bibnamefont
  {Bulgac}},\ }\bibfield  {title} {\enquote {\bibinfo {title} {{Pure quantum
  extension of the semiclassical Boltzmann-Uehling-Uhlenbeck equation}},}\
  }\href {\doibase 10.1103/PhysRevC.105.L021601} {\bibfield  {journal}
  {\bibinfo  {journal} {Phys. Rev. C}\ }\textbf {\bibinfo {volume} {105}},\
  \bibinfo {pages} {L021601} (\bibinfo {year} {2022})}\BibitemShut {NoStop}%
\bibitem [{\citenamefont {Nordheim}(1928)}]{Nordheim:1928}%
  \BibitemOpen
  \bibfield  {author} {\bibinfo {author} {\bibfnamefont {L.~W.}\ \bibnamefont
  {Nordheim}},\ }\bibfield  {title} {\enquote {\bibinfo {title} {{On the
  Kinetic Method in the New Statistics and its Application in the Electron
  Theory of Conductivity}},}\ }\href {\doibase 10.1098/rspa.1928.0126}
  {\bibfield  {journal} {\bibinfo  {journal} {Proc. Roy. Soc. (London)}\
  }\textbf {\bibinfo {volume} {A119}},\ \bibinfo {pages} {689} (\bibinfo {year}
  {1928})}\BibitemShut {NoStop}%
\bibitem [{\citenamefont {Uehling}\ and\ \citenamefont
  {Uhlenbeck}(1933)}]{Uehling:1933}%
  \BibitemOpen
  \bibfield  {author} {\bibinfo {author} {\bibfnamefont {E.~A.}\ \bibnamefont
  {Uehling}}\ and\ \bibinfo {author} {\bibfnamefont {G.~E.}\ \bibnamefont
  {Uhlenbeck}},\ }\bibfield  {title} {\enquote {\bibinfo {title} {{Transport
  Phenomena in Einstein-Bose and Fermi-Dirac Gases. I}},}\ }\href {\doibase
  10.1103/PhysRev.43.552} {\bibfield  {journal} {\bibinfo  {journal} {Phys.
  Rev.}\ }\textbf {\bibinfo {volume} {43}},\ \bibinfo {pages} {552} (\bibinfo
  {year} {1933})}\BibitemShut {NoStop}%
\bibitem [{\citenamefont {Bulgac}\ \emph
  {et~al.}(2024{\natexlab{a}})\citenamefont {Bulgac}, \citenamefont {Kafker},
  \citenamefont {Abdurrahman},\ and\ \citenamefont {Stetcu}}]{Bulgac:2024a}%
  \BibitemOpen
  \bibfield  {author} {\bibinfo {author} {\bibfnamefont {A.}~\bibnamefont
  {Bulgac}}, \bibinfo {author} {\bibfnamefont {M.}~\bibnamefont {Kafker}},
  \bibinfo {author} {\bibfnamefont {I.}~\bibnamefont {Abdurrahman}}, \ and\
  \bibinfo {author} {\bibfnamefont {I.}~\bibnamefont {Stetcu}},\ }\bibfield
  {title} {\enquote {\bibinfo {title} {{Non-Markovian character and
  irreversibility of real-time quantum many-body dynamics}},}\ }\href {\doibase
  10.1103/PhysRevC.109.064617} {\bibfield  {journal} {\bibinfo  {journal}
  {Phys. Rev. C}\ }\textbf {\bibinfo {volume} {109}},\ \bibinfo {pages}
  {064617} (\bibinfo {year} {2024}{\natexlab{a}})}\BibitemShut {NoStop}%
\bibitem [{\citenamefont {Egidy}\ and\ \citenamefont
  {Bucurescu}(2005)}]{Egidy:2005}%
  \BibitemOpen
  \bibfield  {author} {\bibinfo {author} {\bibfnamefont {T.~von}\ \bibnamefont
  {Egidy}}\ and\ \bibinfo {author} {\bibfnamefont {D.}~\bibnamefont
  {Bucurescu}},\ }\bibfield  {title} {\enquote {\bibinfo {title} {Systematics
  of nuclear level density parameters},}\ }\href {\doibase
  10.1103/PhysRevC.72.044311} {\bibfield  {journal} {\bibinfo  {journal} {Phys.
  Rev. C}\ }\textbf {\bibinfo {volume} {72}},\ \bibinfo {pages} {044311}
  (\bibinfo {year} {2005})}\BibitemShut {NoStop}%
\bibitem [{\citenamefont {Goriely}\ \emph {et~al.}(2008)\citenamefont
  {Goriely}, \citenamefont {Hilaire},\ and\ \citenamefont
  {Koning}}]{Goriely:2008}%
  \BibitemOpen
  \bibfield  {author} {\bibinfo {author} {\bibfnamefont {S.}~\bibnamefont
  {Goriely}}, \bibinfo {author} {\bibfnamefont {S.}~\bibnamefont {Hilaire}}, \
  and\ \bibinfo {author} {\bibfnamefont {A.~J.}\ \bibnamefont {Koning}},\
  }\bibfield  {title} {\enquote {\bibinfo {title} {{Improved microscopic
  nuclear level densities within the Hartree-Fock-Bogoliubov plus combinatorial
  method}},}\ }\href {\doibase 10.1103/PhysRevC.78.064307} {\bibfield
  {journal} {\bibinfo  {journal} {Phys. Rev. C}\ }\textbf {\bibinfo {volume}
  {78}},\ \bibinfo {pages} {064307} (\bibinfo {year} {2008})}\BibitemShut
  {NoStop}%
\bibitem [{\citenamefont {Talou}\ and\ \citenamefont {{\it
  others}}(2021)}]{Talou:2021}%
  \BibitemOpen
  \bibfield  {author} {\bibinfo {author} {\bibfnamefont {P.}~\bibnamefont
  {Talou}}\ and\ \bibinfo {author} {\bibnamefont {{\it others}}},\ }\bibfield
  {title} {\enquote {\bibinfo {title} {{Fission fragment decay simulations with
  the CGMF code}},}\ }\href {\doibase
  https://doi.org/10.1016/j.cpc.2021.108087} {\bibfield  {journal} {\bibinfo
  {journal} {Comp. Phys. Com.}\ }\textbf {\bibinfo {volume} {269}},\ \bibinfo
  {pages} {108087} (\bibinfo {year} {2021})}\BibitemShut {NoStop}%
\bibitem [{\citenamefont {Boltzmann}(1872)}]{Boltzmann:1872}%
  \BibitemOpen
  \bibfield  {author} {\bibinfo {author} {\bibfnamefont {L.}~\bibnamefont
  {Boltzmann}},\ }\bibfield  {title} {\enquote {\bibinfo {title} {{Weitere
  Studien {\"u}ber das W{\"a}rmegleichgewicht unter Gasmolek{\"u}len}},}\
  }\href@noop {} {\bibfield  {journal} {\bibinfo  {journal} {Sitzungsberichte
  Akademie der Wissenschaften}\ }\textbf {\bibinfo {volume} {66}},\ \bibinfo
  {pages} {275} (\bibinfo {year} {1872})}\BibitemShut {NoStop}%
\bibitem [{\citenamefont {Jin}\ \emph {et~al.}(2021)\citenamefont {Jin},
  \citenamefont {Roche}, \citenamefont {Stetcu}, \citenamefont {Abdurrahman},\
  and\ \citenamefont {Bulgac}}]{Shi:2021}%
  \BibitemOpen
  \bibfield  {author} {\bibinfo {author} {\bibfnamefont {S.}~\bibnamefont
  {Jin}}, \bibinfo {author} {\bibfnamefont {K.~J.}\ \bibnamefont {Roche}},
  \bibinfo {author} {\bibfnamefont {I.}~\bibnamefont {Stetcu}}, \bibinfo
  {author} {\bibfnamefont {I.}~\bibnamefont {Abdurrahman}}, \ and\ \bibinfo
  {author} {\bibfnamefont {A.}~\bibnamefont {Bulgac}},\ }\bibfield  {title}
  {\enquote {\bibinfo {title} {{The LISE package: solvers for static and
  time-dependent superfluid local density approximation equations in three
  dimentions}},}\ }\href {\doibase 10.1016/j.cpc.2021.108130} {\bibfield
  {journal} {\bibinfo  {journal} {Comp. Phys. Comm.}\ }\textbf {\bibinfo
  {volume} {269}},\ \bibinfo {pages} {108130} (\bibinfo {year}
  {2021})}\BibitemShut {NoStop}%
\bibitem [{\citenamefont {Bulgac}\ \emph {et~al.}(2022)\citenamefont {Bulgac},
  \citenamefont {Abdurrahman},\ and\ \citenamefont
  {Wlaz\l{}owski}}]{Bulgac:2022c}%
  \BibitemOpen
  \bibfield  {author} {\bibinfo {author} {\bibfnamefont {A.}~\bibnamefont
  {Bulgac}}, \bibinfo {author} {\bibfnamefont {I.}~\bibnamefont {Abdurrahman}},
  \ and\ \bibinfo {author} {\bibfnamefont {G.}~\bibnamefont {Wlaz\l{}owski}},\
  }\bibfield  {title} {\enquote {\bibinfo {title} {Sensitivity of
  time-dependent density functional theory to initial conditions},}\ }\href
  {\doibase 10.1103/PhysRevC.105.044601} {\bibfield  {journal} {\bibinfo
  {journal} {Phys. Rev. C}\ }\textbf {\bibinfo {volume} {105}},\ \bibinfo
  {pages} {044601} (\bibinfo {year} {2022})}\BibitemShut {NoStop}%
\bibitem [{\citenamefont {Poincar\'e}(1890)}]{Poincare:1890}%
  \BibitemOpen
  \bibfield  {author} {\bibinfo {author} {\bibfnamefont {H.}~\bibnamefont
  {Poincar\'e}},\ }\bibfield  {title} {\enquote {\bibinfo {title} {{Sur le
  probl\`eme des trois corps et les \'equations de la dynamique}},}\
  }\href@noop {} {\bibfield  {journal} {\bibinfo  {journal} {Acta Mathematica}\
  }\textbf {\bibinfo {volume} {13}},\ \bibinfo {pages} {1} (\bibinfo {year}
  {1890})}\BibitemShut {NoStop}%
\bibitem [{\citenamefont {Neumann}(1929)}]{Neumann:1929}%
  \BibitemOpen
  \bibfield  {author} {\bibinfo {author} {\bibfnamefont {J.~v.}\ \bibnamefont
  {Neumann}},\ }\bibfield  {title} {\enquote {\bibinfo {title} {{Beweis des
  Ergodensatzes und desH-Theorems in der neuen Mechanik}},}\ }\href {\doibase
  10.1007/BF01339852} {\bibfield  {journal} {\bibinfo  {journal} {Zeitschrift
  f{\"u}r Physik}\ }\textbf {\bibinfo {volume} {57}},\ \bibinfo {pages} {30}
  (\bibinfo {year} {1929})}\BibitemShut {NoStop}%
\bibitem [{\citenamefont {von Neumann}(2010)}]{Neumann:2010}%
  \BibitemOpen
  \bibfield  {author} {\bibinfo {author} {\bibfnamefont {J.}~\bibnamefont {von
  Neumann}},\ }\bibfield  {title} {\enquote {\bibinfo {title} {{Proof of the
  ergodic theorem and the H-theorem in quantum mechanics}},}\ }\href {\doibase
  10.1140/epjh/e2010-00008-5} {\bibfield  {journal} {\bibinfo  {journal} {The
  European Physical Journal H}\ }\textbf {\bibinfo {volume} {35}},\ \bibinfo
  {pages} {201} (\bibinfo {year} {2010})}\BibitemShut {NoStop}%
\bibitem [{\citenamefont {Goldstein}\ \emph {et~al.}(2010)\citenamefont
  {Goldstein}, \citenamefont {Lebowitz}, \citenamefont {Tumulka},\ and\
  \citenamefont {Zangh{\`\i}}}]{Goldstein:2010}%
  \BibitemOpen
  \bibfield  {author} {\bibinfo {author} {\bibfnamefont {S.}~\bibnamefont
  {Goldstein}}, \bibinfo {author} {\bibfnamefont {J.~L.}\ \bibnamefont
  {Lebowitz}}, \bibinfo {author} {\bibfnamefont {R.}~\bibnamefont {Tumulka}}, \
  and\ \bibinfo {author} {\bibfnamefont {N.}~\bibnamefont {Zangh{\`\i}}},\
  }\bibfield  {title} {\enquote {\bibinfo {title} {Long-time behavior of
  macroscopic quantum systems},}\ }\href {\doibase 10.1140/epjh/e2010-00007-7}
  {\bibfield  {journal} {\bibinfo  {journal} {The European Physical Journal H}\
  }\textbf {\bibinfo {volume} {35}},\ \bibinfo {pages} {173} (\bibinfo {year}
  {2010})}\BibitemShut {NoStop}%
\bibitem [{\citenamefont {Berry}(1977)}]{Berry:1977}%
  \BibitemOpen
  \bibfield  {author} {\bibinfo {author} {\bibfnamefont {M.~V.}\ \bibnamefont
  {Berry}},\ }\bibfield  {title} {\enquote {\bibinfo {title} {Regular and
  irregular semiclassical wavefunctions},}\ }\href {\doibase
  10.1088/0305-4470/10/12/016} {\bibfield  {journal} {\bibinfo  {journal} {J.
  Phys. A: Mathematical and General}\ }\textbf {\bibinfo {volume} {10}},\
  \bibinfo {pages} {2083} (\bibinfo {year} {1977})}\BibitemShut {NoStop}%
\bibitem [{\citenamefont {Berry}(1991)}]{Berry:1991}%
  \BibitemOpen
  \bibfield  {author} {\bibinfo {author} {\bibfnamefont {M.~V.}\ \bibnamefont
  {Berry}},\ }\enquote {\bibinfo {title} {{Les Houches LII, Chaos and Quantum
  Physics}},}\ \ (\bibinfo  {publisher} {North-Holland, Amsterdam},\ \bibinfo
  {year} {1991})\BibitemShut {NoStop}%
\bibitem [{\citenamefont {Deutsch}(1991)}]{Deutsch:1991}%
  \BibitemOpen
  \bibfield  {author} {\bibinfo {author} {\bibfnamefont {J.~M.}\ \bibnamefont
  {Deutsch}},\ }\bibfield  {title} {\enquote {\bibinfo {title} {Quantum
  statistical mechanics in a closed system},}\ }\href {\doibase
  10.1103/PhysRevA.43.2046} {\bibfield  {journal} {\bibinfo  {journal} {Phys.
  Rev. A}\ }\textbf {\bibinfo {volume} {43}},\ \bibinfo {pages} {2046--2049}
  (\bibinfo {year} {1991})}\BibitemShut {NoStop}%
\bibitem [{\citenamefont {Srednicki}(1994)}]{Srednicki:1994}%
  \BibitemOpen
  \bibfield  {author} {\bibinfo {author} {\bibfnamefont {M.}~\bibnamefont
  {Srednicki}},\ }\bibfield  {title} {\enquote {\bibinfo {title} {Chaos and
  quantum thermalization},}\ }\href {\doibase 10.1103/PhysRevE.50.888}
  {\bibfield  {journal} {\bibinfo  {journal} {Phys. Rev. E}\ }\textbf {\bibinfo
  {volume} {50}},\ \bibinfo {pages} {888} (\bibinfo {year} {1994})}\BibitemShut
  {NoStop}%
\bibitem [{\citenamefont {Rigol}\ and\ \citenamefont
  {Srednicki}(2012)}]{Rigol:2012}%
  \BibitemOpen
  \bibfield  {author} {\bibinfo {author} {\bibfnamefont {M.}~\bibnamefont
  {Rigol}}\ and\ \bibinfo {author} {\bibfnamefont {M.}~\bibnamefont
  {Srednicki}},\ }\bibfield  {title} {\enquote {\bibinfo {title} {{Alternatives
  to Eigenstate Thermalization}},}\ }\href {\doibase
  10.1103/PhysRevLett.108.110601} {\bibfield  {journal} {\bibinfo  {journal}
  {Phys. Rev. Lett.}\ }\textbf {\bibinfo {volume} {108}},\ \bibinfo {pages}
  {110601} (\bibinfo {year} {2012})}\BibitemShut {NoStop}%
\bibitem [{\citenamefont {Bulgac}\ \emph
  {et~al.}(2024{\natexlab{b}})\citenamefont {Bulgac}, \citenamefont {Kafker},
  \citenamefont {Abdurrahman},\ and\ \citenamefont
  {Wlaz\l{}owski}}]{Bulgac:2024}%
  \BibitemOpen
  \bibfield  {author} {\bibinfo {author} {\bibfnamefont {A.}~\bibnamefont
  {Bulgac}}, \bibinfo {author} {\bibfnamefont {M.}~\bibnamefont {Kafker}},
  \bibinfo {author} {\bibfnamefont {I.}~\bibnamefont {Abdurrahman}}, \ and\
  \bibinfo {author} {\bibfnamefont {G.}~\bibnamefont {Wlaz\l{}owski}},\
  }\bibfield  {title} {\enquote {\bibinfo {title} {Quantum turbulence,
  superfluidity, non-markovian dynamics, and wave function thermalization},}\
  }\href {\doibase 10.1103/PhysRevResearch.6.L042003} {\bibfield  {journal}
  {\bibinfo  {journal} {Phys. Rev. Res.}\ }\textbf {\bibinfo {volume} {6}},\
  \bibinfo {pages} {L042003} (\bibinfo {year}
  {2024}{\natexlab{b}})}\BibitemShut {NoStop}%
\bibitem [{\citenamefont {Heller}(1984)}]{Heller:1984}%
  \BibitemOpen
  \bibfield  {author} {\bibinfo {author} {\bibfnamefont {E.~J.}\ \bibnamefont
  {Heller}},\ }\bibfield  {title} {\enquote {\bibinfo {title} {{Bound-State
  Eigenfunctions of Classically Chaotic Hamiltonian Systems: Scars of Periodic
  Orbits}},}\ }\href {\doibase 10.1103/PhysRevLett.53.1515} {\bibfield
  {journal} {\bibinfo  {journal} {Phys. Rev. Lett.}\ }\textbf {\bibinfo
  {volume} {53}},\ \bibinfo {pages} {1515} (\bibinfo {year}
  {1984})}\BibitemShut {NoStop}%
\bibitem [{\citenamefont {Kaplan}\ \emph {et~al.}(2026)\citenamefont {Kaplan},
  \citenamefont {Heller},\ and\ \citenamefont {Keski-Rahkonen}}]{Heller:2026}%
  \BibitemOpen
  \bibfield  {author} {\bibinfo {author} {\bibfnamefont {L.}~\bibnamefont
  {Kaplan}}, \bibinfo {author} {\bibfnamefont {E.}~\bibnamefont {Heller}}, \
  and\ \bibinfo {author} {\bibfnamefont {J.}~\bibnamefont {Keski-Rahkonen}},\
  }\bibfield  {title} {\enquote {\bibinfo {title} {{The Paradoxical Phenomenon
  of Quantum Scarring}},}\ }\href {\doibase 10.1063/pt.5155739a43} {\bibfield
  {journal} {\bibinfo  {journal} {Physics Today}\ }\textbf {\bibinfo {volume}
  {79}},\ \bibinfo {pages} {27} (\bibinfo {year} {2026})}\BibitemShut {NoStop}%
\bibitem [{\citenamefont {Bohigas}\ \emph {et~al.}(1984)\citenamefont
  {Bohigas}, \citenamefont {Giannoni},\ and\ \citenamefont
  {Schmit}}]{Bohigas:1984}%
  \BibitemOpen
  \bibfield  {author} {\bibinfo {author} {\bibfnamefont {O.}~\bibnamefont
  {Bohigas}}, \bibinfo {author} {\bibfnamefont {M.~J.}\ \bibnamefont
  {Giannoni}}, \ and\ \bibinfo {author} {\bibfnamefont {C.}~\bibnamefont
  {Schmit}},\ }\bibfield  {title} {\enquote {\bibinfo {title}
  {{Characterization of Chaotic Quantum Spectra and Universality of Level
  Fluctuation Laws}},}\ }\href {\doibase 10.1103/PhysRevLett.52.1} {\bibfield
  {journal} {\bibinfo  {journal} {Phys. Rev. Lett.}\ }\textbf {\bibinfo
  {volume} {52}},\ \bibinfo {pages} {1} (\bibinfo {year} {1984})}\BibitemShut
  {NoStop}%
\bibitem [{\citenamefont {Bohigas}\ \emph {et~al.}(1993)\citenamefont
  {Bohigas}, \citenamefont {Tomsovic},\ and\ \citenamefont
  {Ullmo}}]{Bohigas:1993}%
  \BibitemOpen
  \bibfield  {author} {\bibinfo {author} {\bibfnamefont {O.}~\bibnamefont
  {Bohigas}}, \bibinfo {author} {\bibfnamefont {S.}~\bibnamefont {Tomsovic}}, \
  and\ \bibinfo {author} {\bibfnamefont {D.}~\bibnamefont {Ullmo}},\ }\bibfield
   {title} {\enquote {\bibinfo {title} {Manifestations of classical phase space
  structures in quantum mechanics},}\ }\href {\doibase
  https://doi.org/10.1016/0370-1573(93)90109-Q} {\bibfield  {journal} {\bibinfo
   {journal} {Physics Reports}\ }\textbf {\bibinfo {volume} {223}},\ \bibinfo
  {pages} {43--133} (\bibinfo {year} {1993})}\BibitemShut {NoStop}%
\bibitem [{\citenamefont {Mehta}(1991)}]{Mehta:1991}%
  \BibitemOpen
  \bibfield  {author} {\bibinfo {author} {\bibfnamefont {M.~L.}\ \bibnamefont
  {Mehta}},\ }\href@noop {} {\emph {\bibinfo {title} {{RANDOM MATRICES and the
  Statistical Theory of Energy Levels}}}}\ (\bibinfo  {publisher} {Academic
  Press, New York},\ \bibinfo {year} {1991})\BibitemShut {NoStop}%
\bibitem [{\citenamefont {Horoi}\ \emph {et~al.}(1995)\citenamefont {Horoi},
  \citenamefont {Zelevinsky},\ and\ \citenamefont {Brown}}]{Horoi:1995}%
  \BibitemOpen
  \bibfield  {author} {\bibinfo {author} {\bibfnamefont {M.}~\bibnamefont
  {Horoi}}, \bibinfo {author} {\bibfnamefont {V.}~\bibnamefont {Zelevinsky}}, \
  and\ \bibinfo {author} {\bibfnamefont {B.~A.}\ \bibnamefont {Brown}},\
  }\bibfield  {title} {\enquote {\bibinfo {title} {{Chaos vs Thermalization in
  the Nuclear Shell Model}},}\ }\href {\doibase 10.1103/PhysRevLett.74.5194}
  {\bibfield  {journal} {\bibinfo  {journal} {Phys. Rev. Lett.}\ }\textbf
  {\bibinfo {volume} {74}},\ \bibinfo {pages} {5194} (\bibinfo {year}
  {1995})}\BibitemShut {NoStop}%
\bibitem [{\citenamefont {Zelevinsky}\ \emph {et~al.}(1996)\citenamefont
  {Zelevinsky}, \citenamefont {Brown}, \citenamefont {F.},\ and\ \citenamefont
  {Horoi}}]{Zelevinsky:1996}%
  \BibitemOpen
  \bibfield  {author} {\bibinfo {author} {\bibfnamefont {V.}~\bibnamefont
  {Zelevinsky}}, \bibinfo {author} {\bibfnamefont {B.~A.}\ \bibnamefont
  {Brown}}, \bibinfo {author} {\bibfnamefont {N.}~\bibnamefont {F.}}, \ and\
  \bibinfo {author} {\bibfnamefont {M.}~\bibnamefont {Horoi}},\ }\bibfield
  {title} {\enquote {\bibinfo {title} {The nuclear shell model as a testing
  ground for many-body quantum chaos},}\ }\href {\doibase
  https://doi.org/10.1016/S0370-1573(96)00007-5} {\bibfield  {journal}
  {\bibinfo  {journal} {Phys. Rep.}\ }\textbf {\bibinfo {volume} {276}},\
  \bibinfo {pages} {85} (\bibinfo {year} {1996})}\BibitemShut {NoStop}%
\bibitem [{\citenamefont {Haake}\ \emph {et~al.}(2018)\citenamefont {Haake},
  \citenamefont {Gnutzmann},\ and\ \citenamefont {Ku\'s}}]{Haake:2018}%
  \BibitemOpen
  \bibfield  {author} {\bibinfo {author} {\bibfnamefont {F.}~\bibnamefont
  {Haake}}, \bibinfo {author} {\bibfnamefont {S.}~\bibnamefont {Gnutzmann}}, \
  and\ \bibinfo {author} {\bibfnamefont {M.}~\bibnamefont {Ku\'s}},\ }\href
  {\doibase 10.1007/978-3-319-97580-1} {\emph {\bibinfo {title} {{Quantum
  Signatures of Chaos}}}}\ (\bibinfo  {publisher} {Springer},\ \bibinfo {year}
  {2018})\BibitemShut {NoStop}%
\bibitem [{\citenamefont {Cohen}\ \emph {et~al.}(1974)\citenamefont {Cohen},
  \citenamefont {Plasil},\ and\ \citenamefont {Swiatecki}}]{Cohen:1974}%
  \BibitemOpen
  \bibfield  {author} {\bibinfo {author} {\bibfnamefont {S.}~\bibnamefont
  {Cohen}}, \bibinfo {author} {\bibfnamefont {F.}~\bibnamefont {Plasil}}, \
  and\ \bibinfo {author} {\bibfnamefont {W.~J.}\ \bibnamefont {Swiatecki}},\
  }\bibfield  {title} {\enquote {\bibinfo {title} {Equilibrium configurations
  of rotating charged or gravitating liquid masses with surface tension. ii},}\
  }\href {\doibase https://doi.org/10.1016/0003-4916(74)90126-2} {\bibfield
  {journal} {\bibinfo  {journal} {Ann. Phys.}\ }\textbf {\bibinfo {volume}
  {82}},\ \bibinfo {pages} {557--596} (\bibinfo {year} {1974})}\BibitemShut
  {NoStop}%
\bibitem [{\citenamefont {Ryssens}\ \emph {et~al.}(2015)\citenamefont
  {Ryssens}, \citenamefont {Heenen},\ and\ \citenamefont
  {Bender}}]{Ryssens:2015}%
  \BibitemOpen
  \bibfield  {author} {\bibinfo {author} {\bibfnamefont {W.}~\bibnamefont
  {Ryssens}}, \bibinfo {author} {\bibfnamefont {P.-H.}\ \bibnamefont {Heenen}},
  \ and\ \bibinfo {author} {\bibfnamefont {M.}~\bibnamefont {Bender}},\
  }\bibfield  {title} {\enquote {\bibinfo {title} {Numerical accuracy of
  mean-field calculations in coordinate space},}\ }\href {\doibase
  10.1103/PhysRevC.92.064318} {\bibfield  {journal} {\bibinfo  {journal} {Phys.
  Rev. C}\ }\textbf {\bibinfo {volume} {92}},\ \bibinfo {pages} {064318}
  (\bibinfo {year} {2015})}\BibitemShut {NoStop}%
\bibitem [{\citenamefont {Abdurrahman}\ \emph {et~al.}(2026)\citenamefont
  {Abdurrahman}, \citenamefont {Kafker}, \citenamefont {Bulgac},\ and\
  \citenamefont {Stetcu}}]{Abdurrahman:2026}%
  \BibitemOpen
  \bibfield  {author} {\bibinfo {author} {\bibfnamefont {I.}~\bibnamefont
  {Abdurrahman}}, \bibinfo {author} {\bibfnamefont {M.}~\bibnamefont {Kafker}},
  \bibinfo {author} {\bibfnamefont {A.}~\bibnamefont {Bulgac}}, \ and\ \bibinfo
  {author} {\bibfnamefont {I.}~\bibnamefont {Stetcu}},\ }\href
  {https://arxiv.org/abs/2607.18511} {\enquote {\bibinfo {title} {Influence of
  the exit channel in $^{235}$u(n,f) and $^{239}$pu(n,f) reactions in
  time-dependent density functional theory},}\ } (\bibinfo {year} {2026}),\
  \Eprint {http://arxiv.org/abs/2607.18511} {arXiv:2607.18511 [nucl-th]}
  \BibitemShut {NoStop}%
\bibitem [{\citenamefont {Bohr}(1956)}]{Bohr:1956}%
  \BibitemOpen
  \bibfield  {author} {\bibinfo {author} {\bibfnamefont {A.}~\bibnamefont
  {Bohr}},\ }in\ \href@noop {} {\emph {\bibinfo {booktitle} {{Proc. Int. Conf.
  Peaceful Uses At. Energy, Geneva, 1955}}}},\ Vol.~\bibinfo {volume} {2}\
  (\bibinfo  {publisher} {United Nations, New York},\ \bibinfo {year} {1956})\
  p.\ \bibinfo {pages} {151}\BibitemShut {NoStop}%
\bibitem [{\citenamefont {Abdurrahman}\ \emph {et~al.}(2024)\citenamefont
  {Abdurrahman}, \citenamefont {Kafker}, \citenamefont {Bulgac},\ and\
  \citenamefont {Stetcu}}]{Abdurrahman:2024}%
  \BibitemOpen
  \bibfield  {author} {\bibinfo {author} {\bibfnamefont {I.}~\bibnamefont
  {Abdurrahman}}, \bibinfo {author} {\bibfnamefont {M.}~\bibnamefont {Kafker}},
  \bibinfo {author} {\bibfnamefont {A.}~\bibnamefont {Bulgac}}, \ and\ \bibinfo
  {author} {\bibfnamefont {I.}~\bibnamefont {Stetcu}},\ }\bibfield  {title}
  {\enquote {\bibinfo {title} {{Neck Rupture and Scission Neutrons in Nuclear
  Fission}},}\ }\href {\doibase 10.1103/PhysRevLett.132.242501} {\bibfield
  {journal} {\bibinfo  {journal} {Phys. Rev. Lett.}\ }\textbf {\bibinfo
  {volume} {132}},\ \bibinfo {pages} {242501} (\bibinfo {year}
  {2024})}\BibitemShut {NoStop}%
\end{thebibliography}%

\clearpage

\end{document}